\documentclass[conference]{IEEEtran}
\IEEEoverridecommandlockouts
\usepackage{cite}
\usepackage{amsmath,amssymb,amsfonts}
\usepackage{algorithmic}
\usepackage{graphicx}
\usepackage{textcomp}
\usepackage{subcaption}
\usepackage{xcolor}
\usepackage{diagbox}
\usepackage{booktabs}
\usepackage{enumitem}
\usepackage[switch, columnwise]{lineno}
\usepackage{blindtext}

\usepackage[utf8]{inputenc}
\usepackage{booktabs,caption}
\usepackage[flushleft]{threeparttable}
\usepackage{ulem}
 
\usepackage{comment}

\def\BibTeX{{\rm B\kern-.05em{\sc i\kern-.025em b}\kern-.08em
    T\kern-.1667em\lower.7ex\hbox{E}\kern-.125emX}}
    
\usepackage{hyperref}
\hypersetup{
     colorlinks = true,
     linkcolor = black,
     anchorcolor = red,
     citecolor = black,
     filecolor = blue,
     urlcolor = black,
}

\begin{document}

\title{BDSP: A Fair Blockchain-enabled Framework for Privacy-Enhanced Enterprise Data Sharing
}

\author{\IEEEauthorblockN{Lam Duc Nguyen, James Hoang, Qin Wang, Qinghua Lu, Sherry Xu,  and Shiping Chen}
\IEEEauthorblockA{\textit{Software Systems Research Group, CSIRO Data61}\\
\textit{Sydney, New South Wales 2015, Australia}}
}        

\maketitle

\begin{abstract}

Across industries, there is an ever-increasing rate of data sharing for collaboration and innovation between organizations and their customers, partners, suppliers, and internal teams. However, many enterprises are restricted from freely sharing data due to regulatory restrictions across different regions,  performance issues in moving large volume data, or requirements to maintain autonomy. In such situations, the enterprise can benefit from the concept of federated learning, in which machine learning models are constructed at various geographic sites.
In this paper, we introduce a general framework, namely $\mathsf{BDSP}$, to share data among enterprises based on Blockchain and federated learning techniques. Specifically, we propose a transparency contribution accounting mechanism to estimate the valuation of data and implement a \textit{proof-of-concept} for further evaluation. The extensive experimental results show that the proposed $\mathsf{BDSP}$ has a competitive performance with higher training accuracy, an increase of over $5\%$, and lower communication overhead, reducing $3$ times, compared to baseline approaches.
\end{abstract}


\section{Introduction}
\label{sec:intro}

Data, an essential strategic resource, has grown significantly in recent years \cite{nguyen2022Blockchain}. Among various industries, the financial sector is leading the way in embracing data as a vital part of the future of business \cite{borgogno2020data}. Enterprises can benefit from pre-trained machine learning (ML) models that are constructed from cross-domain datasets \cite{gaurav2022comprehensive}. However, these enterprises might be unable to readily share data due to regulatory restrictions, large volumes of datasets, or autonomy requirements\cite{briesemeister2019policy}. Generally, the generated and collected datasets contain much sensitive information such as name, gender, and daily operations. Entities that provide these datasets may consequently be reluctant to disclose their private datasets, even though their release would have an essential societal impact.  
For example, according to fraud statistics from the Australia Payments Network \cite{FraudSta89:online}, the overall value of card fraud increased by $5.7\%$, equivalent to $495$ million dollars, in $2021$. Banks can collectively reap the benefits of shared data which are usually forbidden from sharing their transaction datasets \cite{bholat2015big}. In addition, even when several data attributes have been omitted from shared datasets, the aggregated data is still vulnerable to privacy attacks \cite{shokri2017membership}. 

\begin{figure}[t!]
    \centering
    \includegraphics[width=0.9\linewidth]{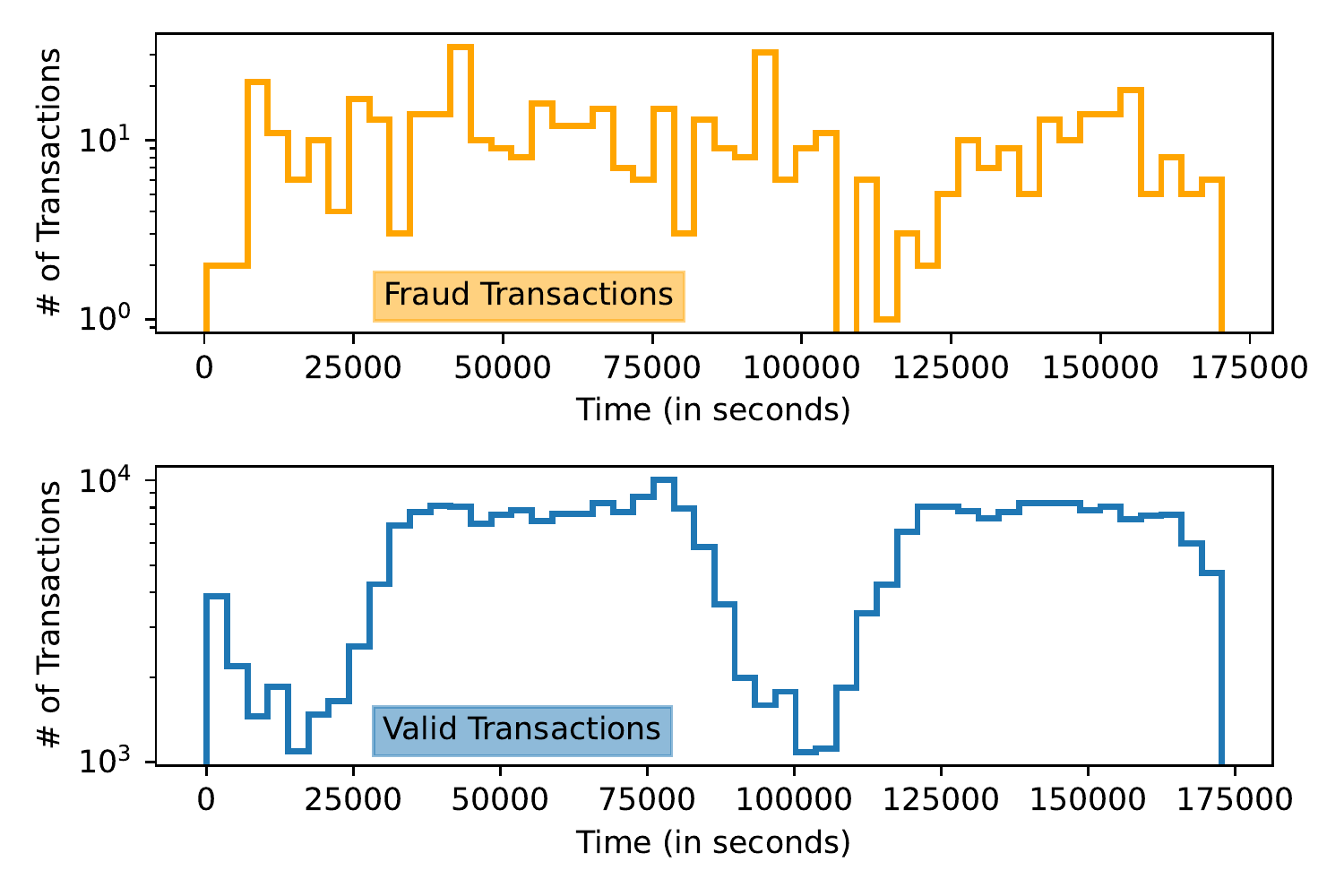}
    \caption{The unbalance of fraud and normal transactions. The data is analyzed from the credit card fraud dataset \cite{CreditCa44:online}.}
    \label{fig:imbalance}
\end{figure}

To address these issues, privacy-preserving techniques are considered as vital approaches for sharing data while guaranteeing the privacy concerns of participants \cite{bernabe2019privacy}. Specifically, there are two main reasons to explain why academia and industrial sectors concentrate on privacy-preserving techniques: i) individuals and organizations are becoming more concerned about the privacy and security of sensitive information, such as names, genders, and races; and ii) the volume of data is increasing exponentially, such as the number of credit card transactions, and the higher requirements for securing sensitive data make it more challenging to anonymize information. Moreover, datasets are highly imbalanced. For example, most of the transactions in every bank are legitimate (see Figure~\ref{fig:imbalance}), leading to bias in constructing ML models.

Various approaches for privacy-preserving approaches have been proposed, such as multi-party computing \cite{goldreich1998secure}, homomorphic encryption \cite{pulido2021privacy}, differential privacy \cite{qu2022personalized}, synthetic data \cite{stadler2022synthetic}, aggregation \cite{yang2021privacy}, and federated learning (FL) \cite{li2021privacy} (a detailed comparison refers to Table~\ref{tab:table-comparision}). In this paper, we focus on privacy-preserving data sharing based on the federated learning method that aims to utilize the locally generated data in a decentralized way without requiring uploading it to a centralized server. Therefore, we can easily maintain the utility of the data even though it is preserved locally. To support organizations smartly using generated data in exchange activities, we introduce an FL-based framework that collaboratively trains FL models using heterogeneous datasets from different companies or banks. 

Previous studies usually deploy conventional FL as a centralized service platform that gathers, shares, and processes raw datasets from their owners for consumers \cite{nguyen2021modeling}. It leads to two vital concerns. First, this strategy exposes this platform as a single point of security risk. 
The malfunctioning platform servers, in particular, pose serious security concerns, including data leakage, inaccurate calculation results, and price manipulation of data. Second, participants or organizations must submit their pre-trained machine learning (ML) models under this strategy.
In particular, researchers leverage Blockchain-based smart contract technology to generate a general ML model by averaging the sum of locally pre-trained ML models from different participants or organizations. 

\begin{figure}[t!]
    \centering
    \includegraphics[width=0.95\linewidth]{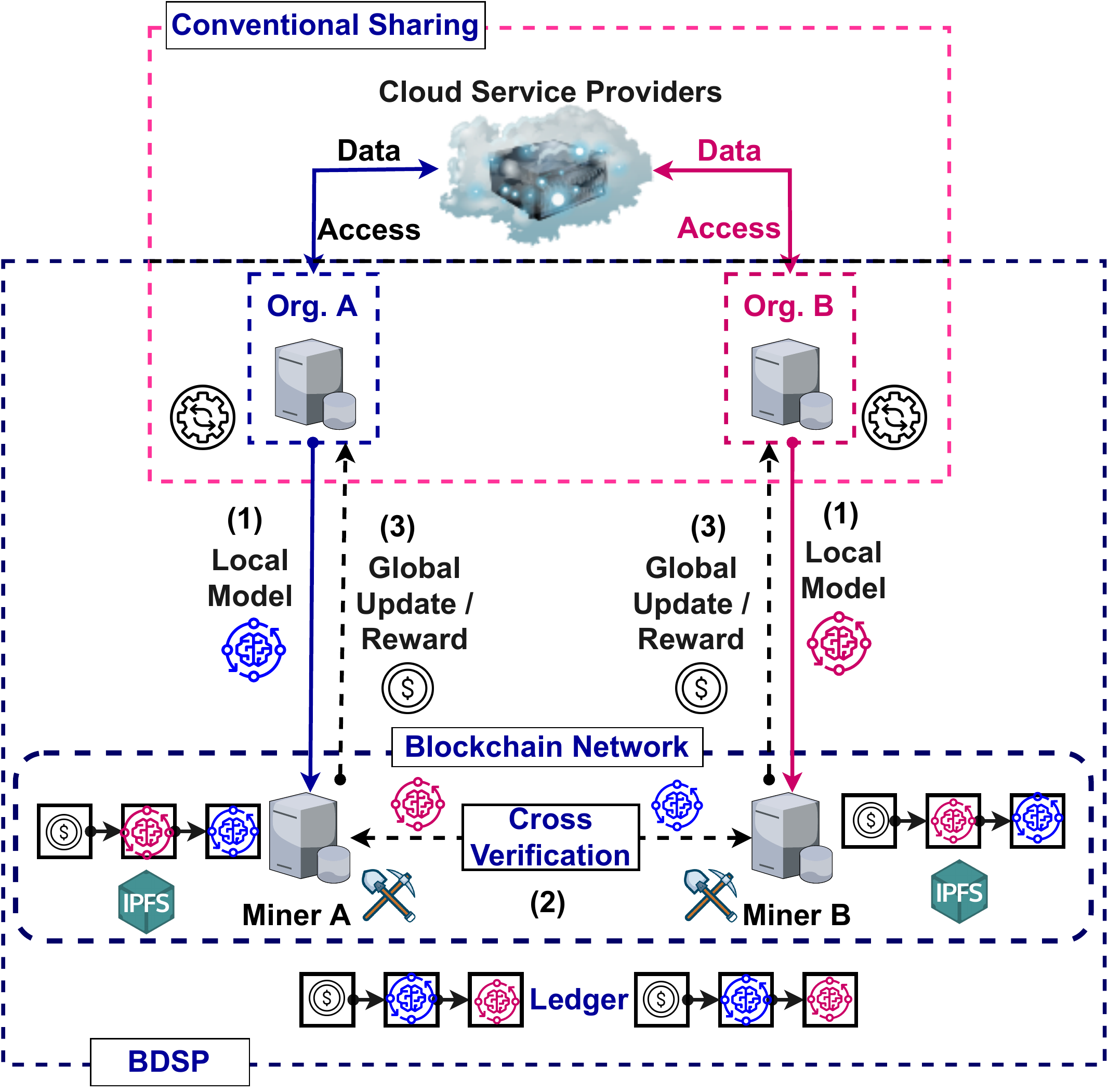}
    \caption{Conventional data sharing and  $\mathsf{BDSP}$: i) the conventional cloud-based service collects data from clients and manages the data access control policies; ii)  $\mathsf{BDSP}$ is designed based on Blockchain and FL, so there is no need to share data among parties to enhance data privacy. }
    \label{fig:comparison}
\end{figure}

Blockchain includes both transaction metadata and contract control policies, which are recorded in a distributed manner among participants \cite{xu2017taxonomy}. Additionally, Blockchain enables the preservation of transactions as immutable records, where each record is spread across multiple participants. Thus, the decentralized nature of Blockchains enables security, as does the use of robust public-key encryption and cryptographic hashes. The advantages of incorporating Blockchains into the exchange of pre-trained ML models in sharing systems include: i) ensuring immutability and transparency for historical ML model sharing records, ii) eliminating the need for third parties, and iii) developing a transparent system for ML model sharing in heterogeneous networks to prevent tampering and injection of fake data from stakeholders.

In this paper, we introduce a general framework named $\mathsf{BDSP}$ that leverages the integration of Blockchain and federated learning (FL) for sharing knowledge among organizations. $\mathsf{BDSP}$ allows organizations in, e.g., finance and hospitals, to collaboratively train FL models without sharing local data.
The general comparison between conventional data sharing and $\mathsf{BDSP}$ is shown in Figure~\ref{fig:comparison}.
$\mathsf{BDSP}$ also focus on the training performance at the client level in order to optimize the performance of the global model. In this paper, the term "clients" stands for organizations. $\mathsf{BDSP}$ addresses the issues of unbalanced datasets among organizations, especially in finances where the number of fraud transactions is minor in comparison with normal transactions. The solution is based on Smote\cite{chawla2002smote}, which helps re-balance the datasets among organizations to achieve a better training collaborated performance. In addition, our proposed framework also provides a mechanism to evaluate the quality of the datasets used to train ML models. In detail, the main \textit{\textbf{contributions}} of this paper are presented as follows: 

\begin{itemize} [leftmargin=*]
    \item[$\triangleright$] First, we systematically analyze the strengths and weaknesses of privacy-preserving techniques for sharing data in the industry at the organizational level. The analysis is used for the chosen data-sharing techniques, used in Blockchain networks, which are demonstrated in section II.  
    
    \item[$\triangleright$] Then, we propose a Blockchain-based framework for sharing data in enterprise environments named $\mathsf{BDSP}$. Instead of sharing raw data, which conflicts with data privacy, we leverage the Federated Learning technique to share information via ML models and keep secret data locally at organizations. In the scope of this research, we consider the deployment of $\mathsf{BDSP}$ in the finance domain, where the finance organizations together share knowledge to detect credit card transactions. 
    
    \item[$\triangleright$] Third, during the training process, we optimize the organization selection based on the contribution from the organization level. In order to do it, we first address the problem of unbalanced datasets among multiple parties and estimate the quality of training data from organizations. For the first point, Smote method is deployed and integrated with $\mathsf{BDSP}$ to rebalance the datasets among organizations before joining the training process. Then, $\mathsf{BDSP}$ provides a federated data valuation mechanism to estimate the quality of training from organizations based on extended Shapley Value. Via extensive experiments, we proved that $\mathsf{BDSP}$ achieves higher accuracy at organization-level around $5\%$ and reduces the convergence time around $3$ times compared to baseline approaches. 
    
    \item[$\triangleright$] Finally, we introduce the design of a proof-of-concept prototype of $\mathsf{BDSP}$ based Hyperledger Fabric v2.4 \cite{androulaki2018hyperledger}, and Pytorch \cite{PyTorch23:online} to run local experiments and evaluate the performance of the proposed framework.
    
\end{itemize}

The rest of the paper is organized as follows. The background knowledge of data sharing techniques is summarized in Section II. Then, the federated data valuation is presented in Section III. The system design is described with detailed components and terminology. Section IV demonstrates the performance analysis. Finally,  we conclude the paper in Section V. 
\section{Data Sharing Solutions and Techniques}
\label{sec:data_techniques}

In this section, we compare the traditional cloud-based data sharing with the Blockchain-based data sharing approach. 
We then analyze privacy-preserving techniques and present some observations for deployment to guarantee the data privacy of involved parties. 

\subsection{Cloud-based Data Sharing} 
With the rapid growth of cloud technology, cloud service providers (CSPs), such as Amazon Web Services\footnote{\url{https://aws.amazon.com/}}, Microsoft Azure\footnote{\url{https://azure.microsoft.com/en-au/}}, and Google Cloud\footnote{\url{https://cloud.google.com/}}, have emerged to offer a range of computing, storage, and networking services \cite{boss2007cloud}. They provide secure, scalable, and cost-effective storage for their customers to transfer, store, and share data in a pay-as-you-go model. 
Customers always have access to their data when they upload it to a cloud environment; hence, customers only need to focus on building businesses based on these data models without taking care of infrastructure \cite{mishra2021analysis}. Using CSPs, however, has some limitations and security concerns. First, due to a lack of trust in sharing data with others, many users and organizations are hesitant to share their sensitive data in the cloud environment \cite{shao2015fine}. Second, the multi-tenancy paradigm of cloud computing, which allows numerous customers to share the same computing resources, i.e., virtual machines, makes it difficult for customers to monitor and manage the use of their data \cite{odun2017cloud}. Finally, customers' data could be leaked, manipulated, and illegally exploited for profit by untrustworthy server operators or CSPs \cite{gupta2019layer}.

\subsection{Blockchain-based Data Sharing}

With the extreme need for sharing data nowadays, it became relevant to explore the use of Blockchain in building decentralized, trusted, and transparent sharing environments \cite{belchior2021survey}. The collected data is formed in Blockchain transactions and recorded in a distributed and redundant manner in a distributed ledger, and each miner verifies transactions. Therefore, it is challenging for malicious parties to attack and manipulate the data for their advantage. Besides, Blockchain technologies  help solve the problem of sharing data in heterogeneous systems. Specifically, organizations usually use different database systems, data formats, data coding, and access interfaces adopted by various government sectors. However, these sectors often lack support for a unified data model and interfaces.

For instance, the authors in \cite{jaiman2020consent} developed a dynamic consent model for sharing medical data based on Blockchain. In this system, each data provider decides on data use conditions that data requesters must follow. Such a model ensures that individual consent is secured and that all platform participants are held accountable.
%
The authors in \cite{nguyen2021modeling} introduced a model for data sharing based on smart contracts via a Narrowband Internet of Things (NB-IoT) system, aiming to support massive environmental sensing. The authors analyzed the communication efficiency and three data-sharing protocols in terms of latency and energy consumption over NB-IoT communication. The proposed model could be considered as a benchmark for sharing IoT data in resource-constrained networks. However, data recorded in Blockchain is immutable, so this could be a challenge to adapt Blockchain as a part of industrial solutions. 
The authors in \cite{nguyen2021marketplace} proposed an ML model-sharing platform based on Blockchain smart contracts. In general, the authors integrated Smart Contracts and Federated Learning for sharing models, and estimate the model price based on the valuation of datasets. 

\subsection{Privacy-Preserving Techniques for Data Sharing}
\subsubsection{Secure Multi-party Computation}
This approach presents a technology paradigm based on cryptography that enables privacy protection in sharing data among multiple involved participants \cite{yao1982protocols}. Theoretically, secure multi-party computation (SMPC) enables a balance of tension between sharing data to create value and securing data as a competitive advantage\cite{du2001secure}. Compared to traditional cryptography, in which cryptography assures security and integrity of communication or storage, and the adversary is outside the system of participants, e.g., an eavesdropper on the sender and receiver, cryptography in SMPC enables the protection of the participant's privacy from each other\cite{yang2020block}. However, SMPC is challenging to build and has complex computation requirements. For the deployment, this approach requires establishing separate computation and process servlets.

\begin{table}[!h]
\centering
\caption{Quantitative comparison of privacy-preserving data sharing strategies 
}
\begin{threeparttable}
\resizebox{\linewidth}{!}{
\begin{tabular}{l c c c c c c} 

\toprule
\multicolumn{1}{c}{\diagbox{\textbf{PPT}}{\textbf{Attributes}}} & \begin{tabular}[c]{@{}c@{}}\textbf{Privacy}\end{tabular} & \textbf{Size~} & \textbf{Speed} & \textbf{Data Utility} & \textbf{Deployment}  \\ 
\midrule
\textit{Secure Multi-party Computing}    
&  $\star$$\star$$\star$$\star$$\star$     
& $\star$$\star$      
& $\star$$\star$    
& $\star$$\star$$\star$$\star$         
& $\star$ \\ 

\textit{Homomorphic Encryption}         
& $\star$$\star$$\star$$\star$$\star$     
& $\star$$\star$   
& $\star$$\star$    
& $\star$$\star$$\star$$\star$              
& $\star$  \\ 

\textit{Differential Privacy}           
& $\star$$\star$$\star$$\star$    
&  $\star$$\star$$\star$$\star$        
&  $\star$$\star$$\star$$\star$  
& $\star$$\star$$\star$               
& $\star$$\star$$\star$  \\ 
\textit{Synthetic Data}                 
& $\star$$\star$$\star$$\star$$\star$ 
&  $\star$$\star$$\star$       
& $\star$$\star$$\star$        
& $\star$$\star$$\star$              
& $\star$$\star$$\star$ \\ 

\textit{Federated Learning}             
& $\star$$\star$$\star$$\star$$\star$   
& $\star$$\star$$\star$       
& $\star$$\star$$\star$       
& $\star$          
& $\star$$\star$ \\ 
\bottomrule
\end{tabular}
}
\end{threeparttable}
\label{tab:table-comparision}
\end{table}

\subsubsection{Homomorphic Encryption}
With homomorphic encryption (HE)\cite{acar2018survey}, organizations can configure a higher standard of data security and privacy without breaking processes or existing application functionalities. HE is usually used in cloud workload protection, aggregate analysis, and automation. Theoretically, for a given HE function $E$ with respect to a function $f$, the encrypted of $f$ can be computed by computing a function $g$, over encrypted variables of $X = \{x_1, x_2, ... x_n\}$.
\begin{equation}
    E(f(x_1, x_2, ..., x_n)) \equiv g(E(x_1), E(x_2), ..., E(x_n))
\end{equation}
Each participant can encrypt his private $x_i$ and transmit $E(x_i)$ to another. Next, that party calculates the function $g$ on encrypted data via homomorphism of the encryption method, and the party gets the encrypted value of function $f$. However, it is still a challenge to apply HE in real applications \cite{li2021privacy} as HE requires either application modifications or specialized client-server applications to make it work.  

\subsubsection{Differential Privacy}
Another possible solution to enhance data when making sensitive private data securely available is differential privacy (DP). The definition of DP is defined as follows: a randomized algorithm $\mathcal{A}: \mathcal{D} \rightarrow \mathcal{R}$ with domain $\mathcal{D}$ and range $\mathcal{R}$ is $(\epsilon, \delta)$ - differential privacy if for any two adjacent training dataset $D_1, D_2 \subseteq \mathcal{D}$, in which the data points in these two datasets are difference, and any subset of output $S\in \mathcal{R}$, satisfies the condition: 
\begin{equation}
    Pr[\mathcal{A}(D_1) \in S ] \leq e^\epsilon Pr[\mathcal{A}(D_2) \in S] + \delta,
\end{equation}
where $\epsilon$ and $\delta$ are called privacy budget and failure rate, respectively. A smaller $\epsilon$, a stronger privacy guarantee. 
However, DP also has some serious flaws \cite{domingo2021limits}. Some DP queries leak a small amount of data, hence if an attacker is able to repeat similar queries, the total loss could be catastrophic.

\subsubsection{Synthetic Data}

Synthetic data \cite{wang2019learning} is a type of artificial data that mimics realistic observations and is used to train AI and ML models when there is a lack of actual data. The synthetic data looks like the original datasets, shares the same schema, and attempts to maintain properties of the original ones, e.g., features and correlations. Even though using synthetic data brings advantages, it still has some drawbacks, such as failing to represent real-world events and containing bias \cite{abay2018privacy}. In addition, the synthetic data approach does not allow participants to combine data across datasets from the same individual. 

\subsubsection{Federated Learning}


Federated learning (FL) \cite{bonawitz2019towards} is defined as an ML technique, in which multiple entities, called clients, collaborate in solving a machine learning problem under the coordination of a central aggregator or service provider without clients having actually to exchange their data \cite{nguyen2021marketplace}. Different from traditional machine learning, FL enables learning a shared model collaboratively. The shared model is further trained under the coordination of a central server by using a dataset distributed on the participating devices and defaulting to privacy. 
We describe the details of FL in Section~\ref{sec:federated}. 

\textbf{Observation:} According to the analysis, each technique has different advantages and disadvantages. Depending on specific applications and domains, we can choose one of the above strategies or combine these methods to satisfy requirements regarding privacy concerns and system resources. Table~\ref{tab:table-comparision} describes a quantitative comparison of privacy-preserving techniques for data sharing.
Both secure multi-party computation (SMPC) and homomorphic encryption (HE) require the use of advanced cryptographic techniques and may not be feasible for all enterprises. However, for those that are able to implement them, they can provide a strong foundation for privacy-preserving data sharing.
In general, SMPC and HE protocols provide the strongest guarantees of privacy, but they may be more complex to implement and may not be suitable for all applications. Differential privacy is a more flexible approach, but it may not provide the same level of privacy protection as the other techniques.
In the scope of this research, we combine synthetic data and federated learning for the system design. Besides, the consent of credit card users should use data initially generated by their activities.
\section{Federated Data Valuation}
\label{sec:federated}

This section describes the concept of federated learning and the algorithm used to estimate the valuation of data based on Shapley value \cite{nguyen2021marketplace, ghorbani2019data}.

\begin{figure}
    \centering
    \includegraphics[width=0.9\linewidth]{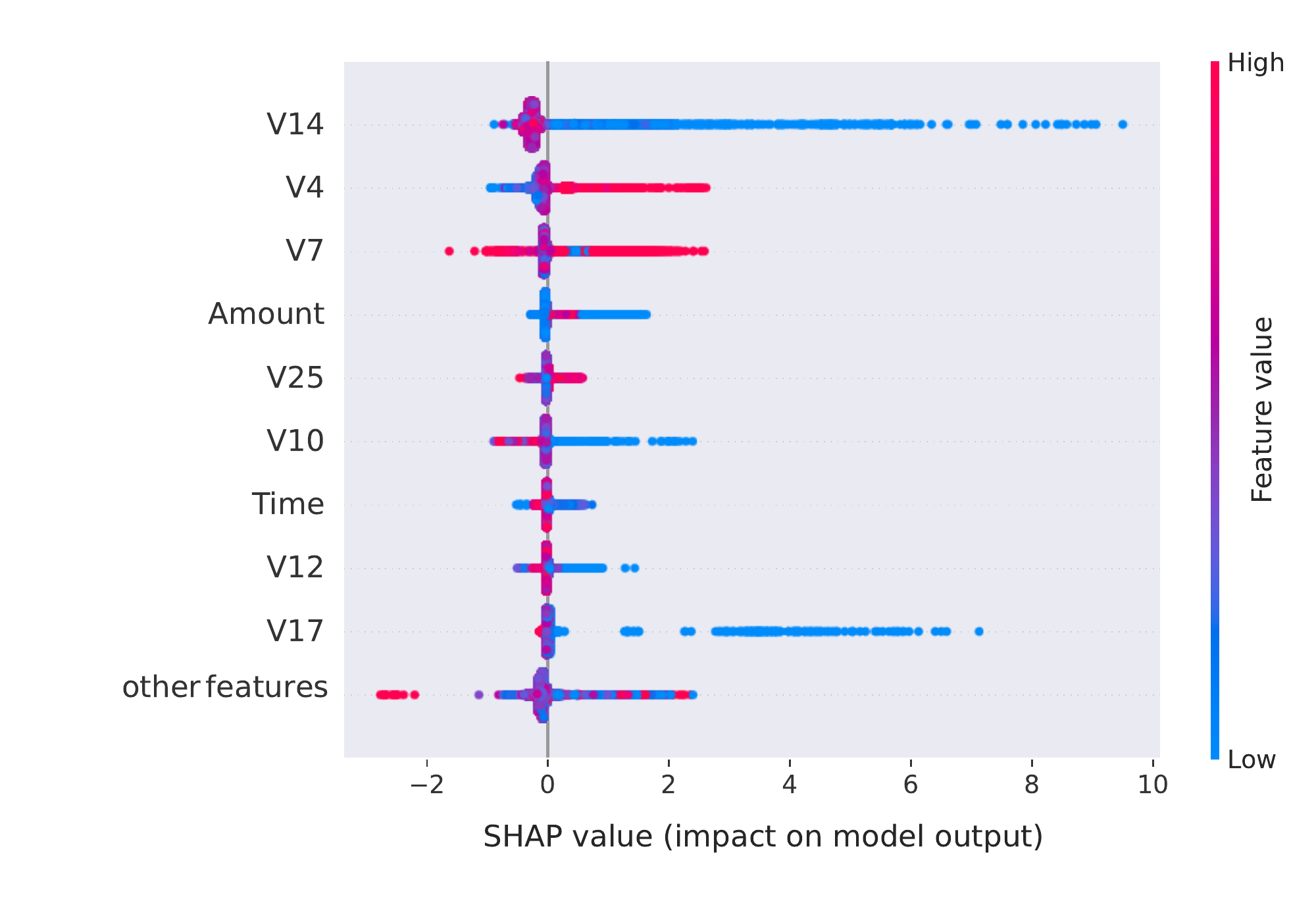}
    \caption{The imbalance in the importance of features analyzed from the dataset. For example, the feature V14 shows the highest score and impact on the output of the training model.}
    \label{fig:feature_important}
\end{figure}

\subsection{Federated Learning}
Suppose that the organization-data owner $i$-th has a training dataset $D_i$. In federated learning (FL), this translates to finding an optimal global model parameter $w \in \mathbb{R}^d$ that minimizes the empirical risk on all of the  distributed training data samples, the clients collaborate to solve the distributed optimization problem as follows:
\begin{equation}
    \underset{w \in \mathbb{R}^\textrm{d}}{\text{min}}F(w) := \frac{1}{N} \sum_{i=1}^{N} F_i(w)
\label{eq:objectivefunction}
\end{equation}
where $N$ is the number of data providers. The local objective for clients $i$ is as follows:
\begin{equation}
    F_i(w) := L(w; D_i),
\end{equation}
where $L$ is a loss function \cite{chu2004bayesian}. We assume that the configuration ensures that if two FL clients have identical local datasets, then they have same local models, e.g, $D_i = D_j$, so $F_i = F_j$. In order to solve the optimization problem of (\ref{eq:objectivefunction}), we leverage the federated averaging (FedAvg) algorithm \cite{bonawitz2019towards}, which is a widely used FL. FedAvg performs stochastic gradient descent (SGD)\cite{malinovskiy2020local} parallelly on a randomly sampled subset of clients and submits local model updates to a central server in each round. Specifically, $I=\{1,...,N\}$ is the set of FL clients, in each training round $t$, FedAvg performs as follows: 
\begin{enumerate}
    \item First, the centralized server broadcasts a last global model $w^t$ to all of FL clients, say data providers. 
    \item Then, all data provider $i$ updates its local model by configuring  $w_i^t = w^t$ for all FL client $i$. 
    \begin{equation}
        w_i^{t_1} = w_i - \eta^t \nabla F_i(w_i^t), 
    \end{equation}
    where learning rate $n^t$ is used in round $t$-th. The default value of $n^t=0.1$
    \item Based on the predefined client selection scheduling algorithm, a subset $I_t \subseteq I$ of FL clients is selected. FedAvg algorithm randomly selects clients for local updates.
    \item Next, the central server aggregates the submitted local models to generate a new global model
    \begin{equation}
        w^{t+1} = \frac{1}{|I_t|} \sum_{i\in I_t} w^{t+1}
    \end{equation}
\end{enumerate}

The FL training process is iterated until the global loss function converges or archives a desirable test accuracy.

\subsection{Re-blancing Datasets}


The unbalance of a data class is a common issue in classification tasks in which one class output is considered a minority class, and has fewer data samples than other output classes in a binary classification dataset \cite{kaur2019systematic}. The imbalance dataset issue affects the performance of model training and causes it to favor the majority class while missing the minority class. For example, Fig~.\ref{fig:feature_important} shows the imbalance in the importance of features. In this paper, we deploy the SMOTE method \cite{chawla2002smote} to rebalance the credit card datasets among finance organizations. 

\begin{equation} 
\forall x_{i} \in D_{min};\quad \text {return} \hspace {1mm}\{knn\}
\end{equation}

Then the SMOTE method randomly selects one of the neighbors to perform the linear interpolation. The newly generated samples from SMOTE are as follows: 
\begin{equation} 
    x_{new} = x_{i} + (x_{i} - x_{ij}){RAND}[{0,1}],
\end{equation}
where $x_{new}$ is the newly generated synthetic sample added to the training dataset. 


\subsection{Federated Data Valuation}
There are several methods to evaluate the valuation of training data in FL. For example, a permutation-based method \cite{koh2017understanding} for clarifying training data points for a given prediction, a reinforcement learning-based method to study the contribution of each data point towards the predictor model \cite{yoon2020data}, and a Shapley-based data valuation \cite{nguyen2021marketplace}. 
For proof-of-concept purposes, we exploit the classic Shapley-based data valuation method for evaluating the valuation of datasets. 

In FL training, data valuation targets finding datasets that are relevant and important to the learning results. The importance of quality datasets is reflected in the improvements and performance of the final global models. Denote that at round $t$-th of training, for $1 \leqslant t \leqslant T $, given the utility function $U_t: 2^I \rightarrow R$ for any subset of FL clients $S \subset I$, $U(S)$ return a utility score of the global model that collaboratively trained by FL clients in set $S$. At round $t$, the utility created by FL clients is defined as follows: 
\begin{equation}
    U^t(S) := u^t \big(\frac{1}{|S|} \sum_{k\in S} w_k^{t+1} \big ),
\end{equation}
where $u_t: R^n \rightarrow R, u^t(w) = L(w^t, D_s) - L(w, D_s)$ is the utility function each round, $D_s$ is the test data located in the central server. The $u_t(w)$ presents the global performance in each round $t$ with the parameter $w$. The Shapley-based strategy has three main requirements as follows: 

\begin{itemize}
    \item \textit{Symmetry}: For all $S \subseteq I \setminus \{ I, j\}$, if users $i$, and $j$ are interchangeable, and $U(S \cup {i}) = U(S\cup {j} )$, then data valuation metric $v_i = v_j $, where 
    \begin{equation}
         v_i = c \sum_{S \subseteq I \setminus \{i\}}  \frac{1}{\begin{pmatrix} N-1\\ |S| \end{pmatrix}} \big [ U(S\cup \{i\}) - U(S) \big ] .
    \end{equation}
   
    \item \textit{Dummy participants}: Participant $i$ is a dummy participant if the amount data of $i$ contribute to coalition is exactly the amount $i$ can achieve alone, e.g, $\forall S$, $i \notin S$, $U(S \cup \{i\} - U(S) = U(i)$. 

    \item \textit{Additivity}:  For any two $U_i$ and $U_j$, we have for any participant $i$, we have $v_i(N,U_i + U_j)=v_i(N,U_i) + v_i(N,U_j)$, where the game $(N,U_i + U_j)$ is defined by $(U_i + U_j)(S) = U_i(S) + U_j(S)$ for every coalition $S$.
\end{itemize}

In this research, we also exploit the contribution-based method \cite{pandey2022contribution} for organization selection with modified Truncated Monte-Carlo strategy\cite{ghorbani2019data} within random organization selection rounds. In fact, plain FL usually ignores the performance from client-level because of a small amount of data or limited resource capacities. Therefore, the random client selection procedure may increase costs in terms of overhead. However, at the organization-level, when the volume of the dataset is bigger, we need to address the performance of each organization that contributes to the training process. The detail is presented in \cite{pandey2022contribution}.

\section{System Design}
\begin{figure}[t!]
    \centering
    \includegraphics[width=0.99\linewidth]{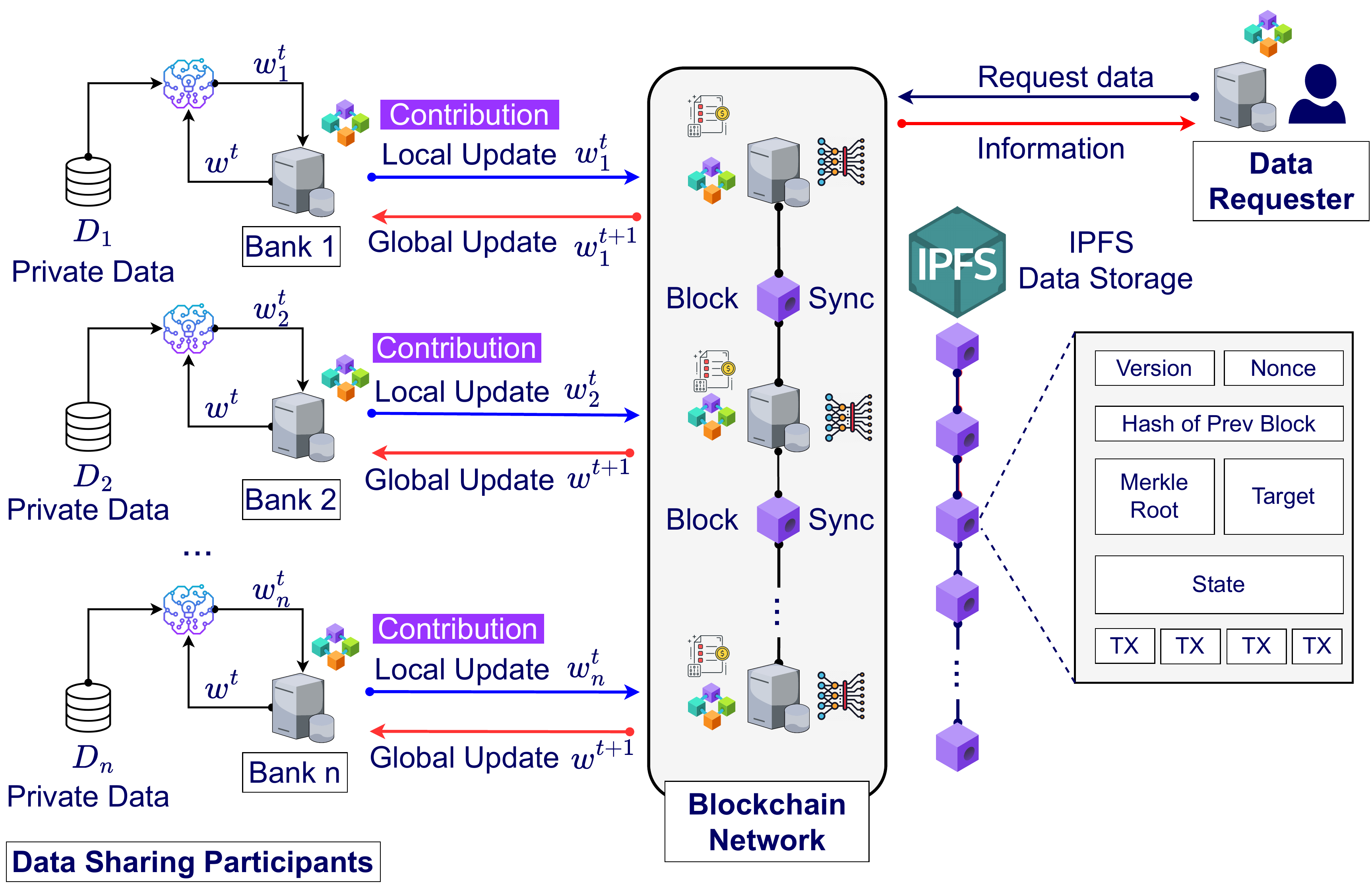}
    \caption{$\mathsf{BDSP}$ general system architecture includes four components, namely Blockchain network, data provider, data request, and external services.}
    \label{fig:architecture}
\end{figure}

In this section, we describe the general architecture of the $\mathsf{BDSP}$ platform (cf. Figure~\ref{fig:architecture}) and the communication workflow of the proposed system.

\subsection{General Architecture}
The $\mathsf{BDSP}$ system includes four main components, such as, distributed ledger, data providers, data requesters, and external services. We present them as follows:

\noindent\hangindent 1em\textit{Distributed ledger.} This component includes all modules to build various features of Blockchain technologies, such as consensus, smart contracts, data authorization, identity management, and peer-to-peer (P2P) communication. These components must ensure that every change to the ledger is reflected in all copies in seconds or minutes and provide mechanisms for the secure storage of the data created by participants and parameter configurations. 

\noindent\hangindent 1em\textit{Data providers.} The organizations would like to share their valuable data with authorized parties through a Blockchain-based data-sharing platform. These data providers have full permission to control all access to their own data. They are responsible for defining access policies, including information about encryption and metadata. 

\noindent\hangindent 1em\textit{Data requester.} The organizations aim to train their model to predict fraudulent transactions. However, because of the lack of datasets, they need to collaborate with other organizations to achieve better results, e.g., improve accuracy and reduce latency. 

\noindent\hangindent 1em\textit{External services.} 

Every day, financial institutions have to deal with a massive volume of transactions. External infrastructure that operates on the edge may be incorporated to provide external services, such as storage and computing. 
The Interplanetary File System (IPFS) \cite{chen2017improved}, for example, is a distributed file storage system capable of storing data generated by transactions and control services. Moreover, it also returns a hash to the ledger based on the content of the data. Because the ledger cannot handle and store the massive amount of generated transactions, the IPFS service is critical. IPFS's default configuration connects it to the global distribution network. We prefer connecting to a private IPFS network over a public one in cases involving privacy and confidentiality.

\subsection{Communication Workflow}
In this section, we describe the communication workflow of the $\mathsf{BDSP}$ system. Figure~\ref{fig:architecture} the details of our proposed system. 
We assume a scenario in which a finance organization, e.g., a bank, agrees to share credit card transactions for ML training purposes to build a higher accuracy system and efficiency prediction to prevent fraudulent transactions. Each organization $B_i$, denoted as \texttt{Bank} in Figure~\ref{fig:architecture}, has its own ML model and dataset. We present the process of the BDSP system as follows:

\noindent\hangindent 1em\textit{Step 1 - Model initialization:} 
%
The organization $B_i$ initiates a model $w_i$ that needs to be trained and shares the model with other organizations. Note that these other organizations also have their data and are willing to train the model for $B_i$. 

\noindent\hangindent 1em\textit{Step 2 - Local training:} 
%
The involved organization $B_j$ first downloads the initial model $w_i$. Then $B_j$ trains the model $w_i$ with its local dataset $D_j$, which is kept private from organizations. 

\noindent\hangindent 1em\textit{Step 3 - Cross verification of the local models:} 
%
After receiving the local model published by organizations in the format of transactions, the Blockchain miner places this model in newly generated blocks and broadcasts it to other Blockchain miners in the network. 
Note that, only selected organizations can submit trained local models to the aggregator. These organizations are selected based on the contribution-based device selection process as described in \cite{pandey2022contribution}.
Next, until other Blockchain miners receive the broadcasted blocks, which include the local models of clients, they verify the accuracy of the local models and allocate the models to the newly generated blocks. During this process, we broadcast all aggregated models to all the Blockchain miners in the system. These Blockchain miners further compare the consistency and accuracy among the aggregated models. We then select the most popular model as the correct global model. Blockchain miners record this model and the contributions of the organizations into the distributed ledger via smart contract features. Otherwise, we consider the rest of the global models to be faulty updates.

\noindent\hangindent 1em\textit{Step 4 - Model aggregation:} Next, the organization $B_j$ is randomly assigned with a Blockchain validator to upload its trained local model $w_j$ to aggregate. 

\noindent\hangindent 1em\textit{Step 5 - Finalization:} These above steps are repeated until achieving a certain accuracy or after a pre-defined number of training rounds. 

\begin{table}[!h]
\centering
\caption{Parameter settings}
\label{tab:parameter-setting}
\begin{tabular}{lc}
\toprule
\multicolumn{1}{c}{\textbf{Parameter}} & \textbf{Value} \\
\midrule
Total Number of Organizations                           & 30             \\
Number of selected clients per round           & 10             \\
Training Rounds                        & 100            \\
Learning Rate                          & 0.01           \\
Number of Epochs                       & 5,10,15        \\
Batch Size                             & 16,32,64       \\
Weight Decay                           & 0.001          \\
Model                                  & CNN            \\
Optimizer                              & SGD           \\
Performance Metrics                    & Acc, Loss, F1, Precision \\
\bottomrule
\end{tabular}
\end{table}

\begin{table*}[!t]
\centering
\caption{System experimental configuration 
}
\label{tab:exp_config}
\begin{tabular}{lllll}
\toprule
\multicolumn{1}{c}{\textbf{System Component}} &
  \multicolumn{1}{c}{\textbf{Version}} &
  \multicolumn{1}{c}{\textbf{CPU}} &
  \multicolumn{1}{c}{\textbf{RAM}} &
  \multicolumn{1}{c}{\textbf{Disk}} \\
\midrule
Hyperledger Fabric  & v.24   & Intel(R) Xeon(R) Gold 6248 CPU$^{\star}$@2.50GHz & 16GB & 100GB \\
Hyperledger Caliper & v0.5.0 & Intel(R) Xeon(R) Gold 6248 CPU@2.50GHz & 16GB & 100GB \\
HLF Peer            & v.24   & Intel(R) Xeon(R) Gold 6248 CPU@2.50GHz & 16GB & 100GB \\
FL Library        & Pytorch & Intel(R) Xeon(R) Gold 6248 CPU@2.50GHz & 16GB & 100GB \\
\bottomrule
\end{tabular}

\end{table*}

\begin{figure*}[t!]
     \centering
     \begin{subfigure}[b]{0.24\textwidth}
         \centering
         \includegraphics[width=\textwidth]{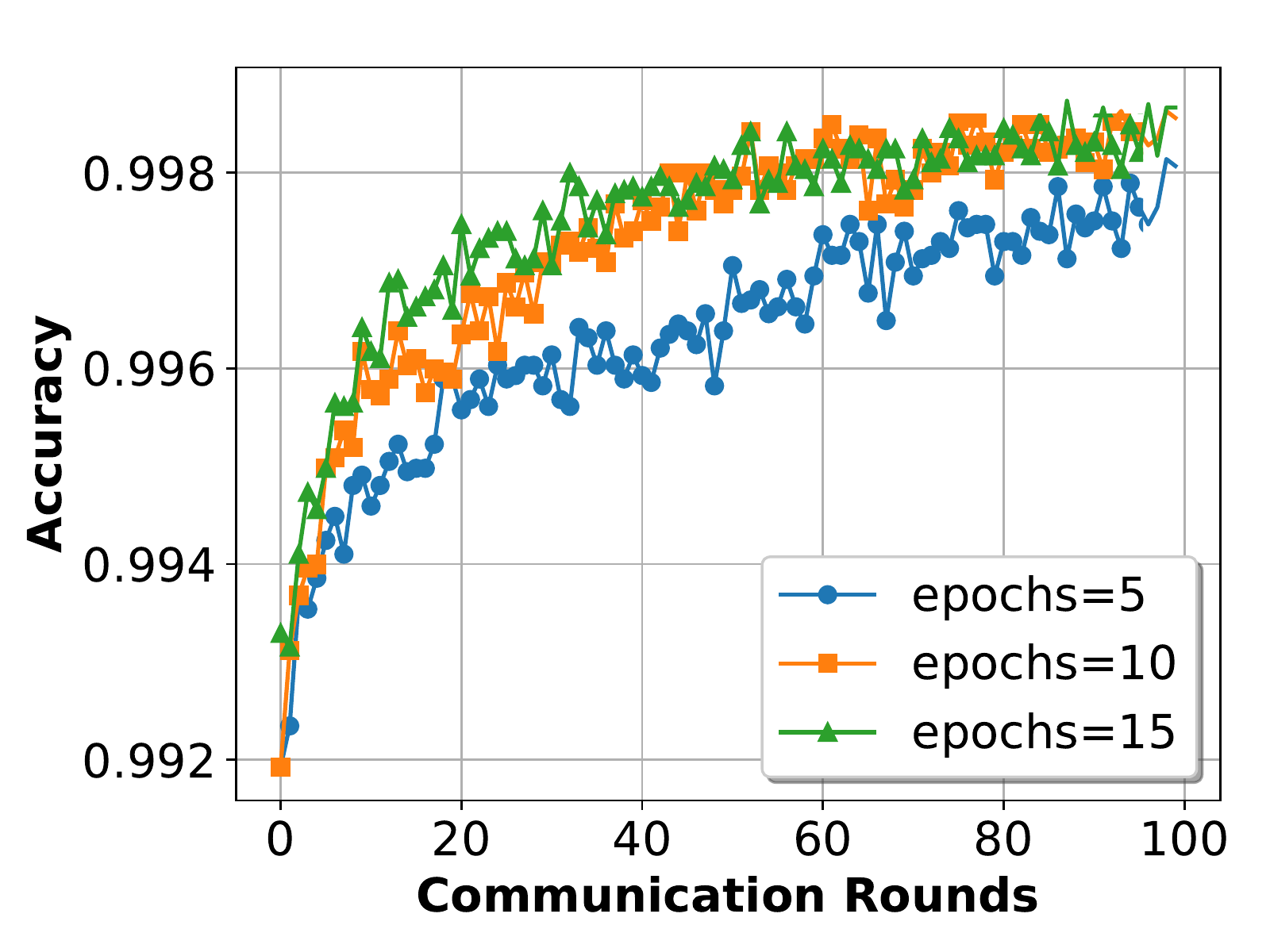}
         \caption{Test Accuracy}
         \label{fig:acc-epoch}
     \end{subfigure}
     \begin{subfigure}[b]{0.24\textwidth}
         \centering
         \includegraphics[width=\textwidth]{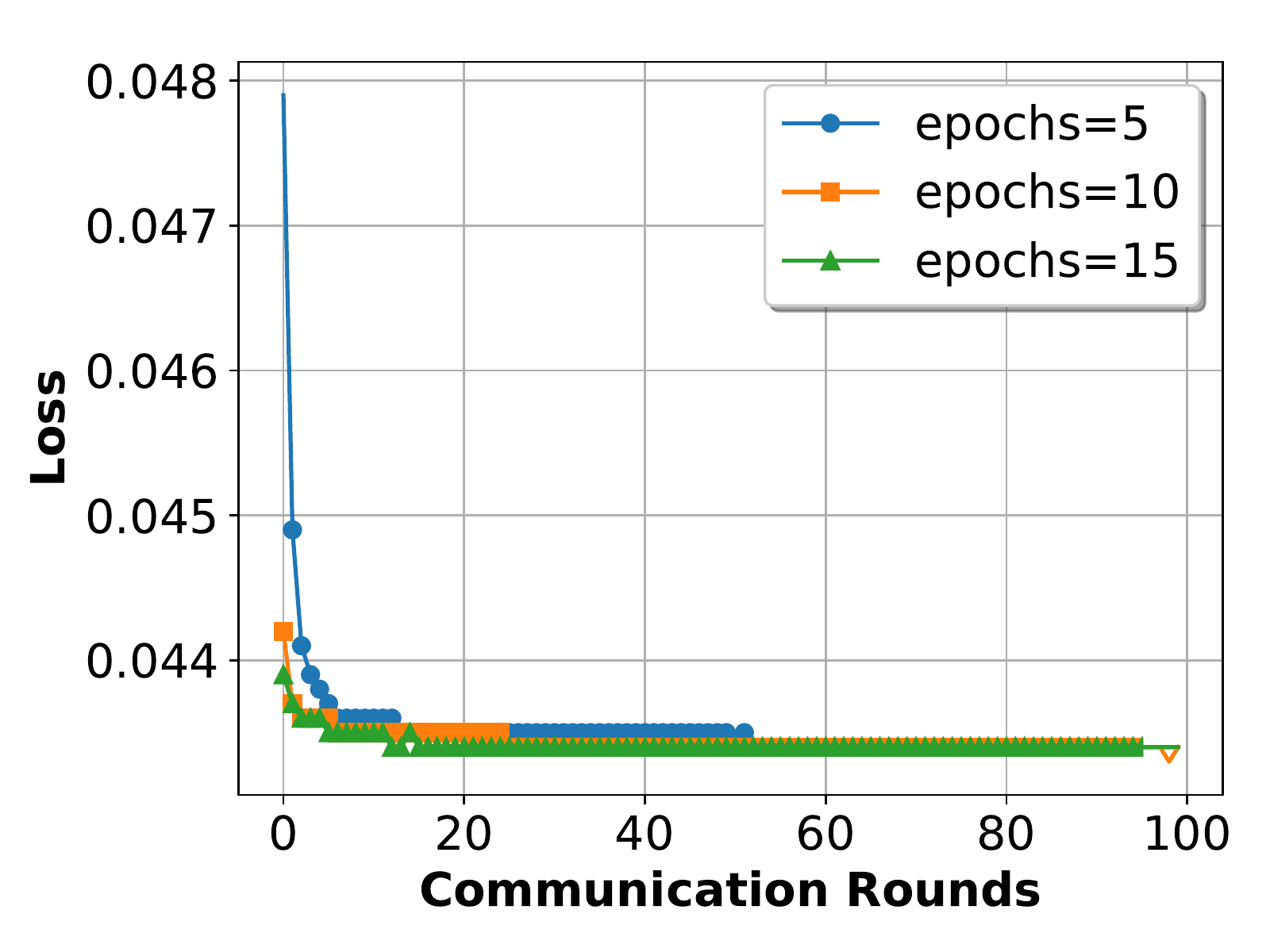}
         \caption{Test Loss}
         \label{fig:loss-epoch}
     \end{subfigure}
     \begin{subfigure}[b]{0.24\textwidth}
         \centering
         \includegraphics[width=\textwidth]{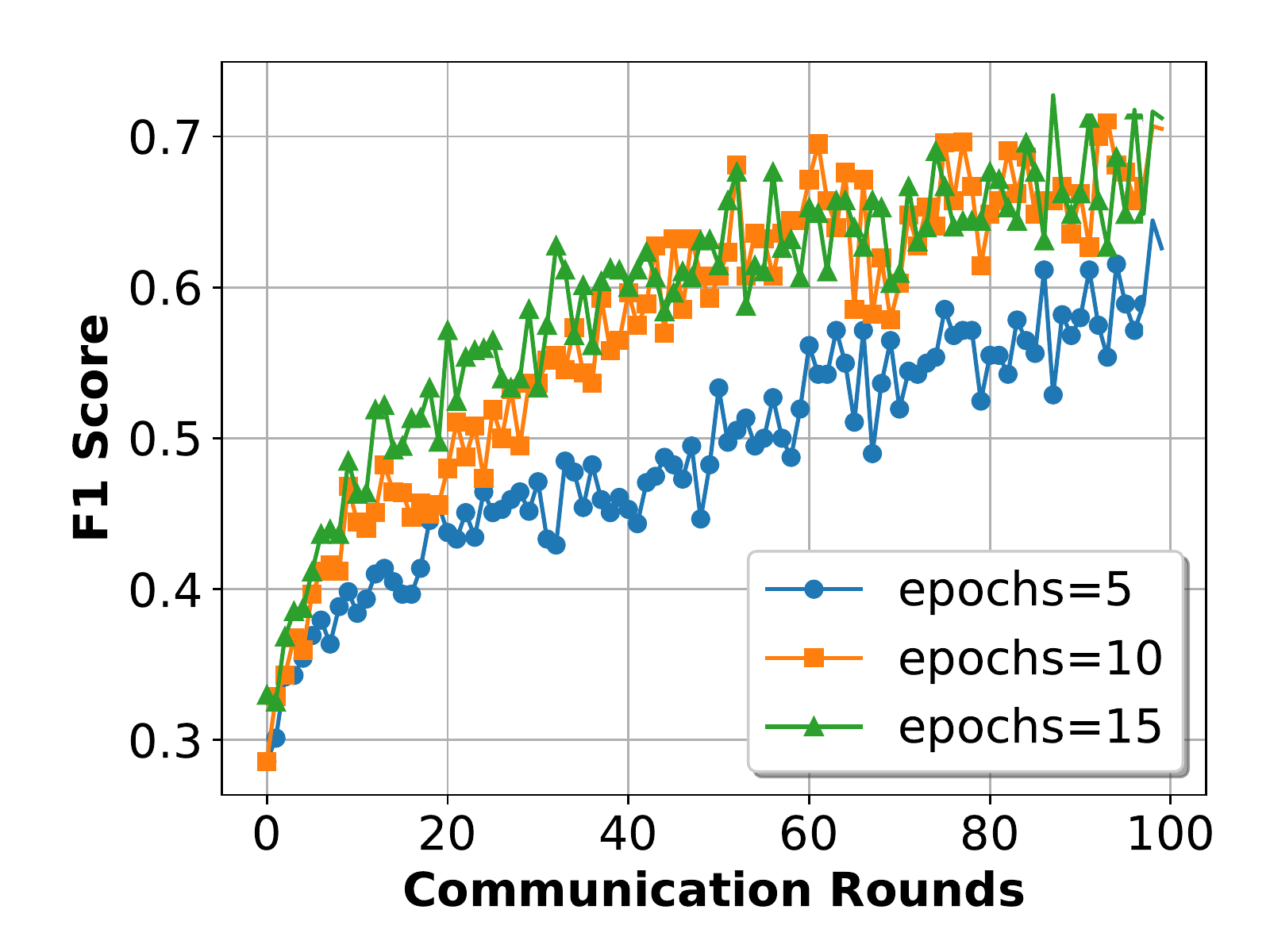}
         \caption{F1 Score}
         \label{fig:f1-epochs}
     \end{subfigure}
     \begin{subfigure}[b]{0.24\textwidth}
         \centering
         \includegraphics[width=\textwidth]{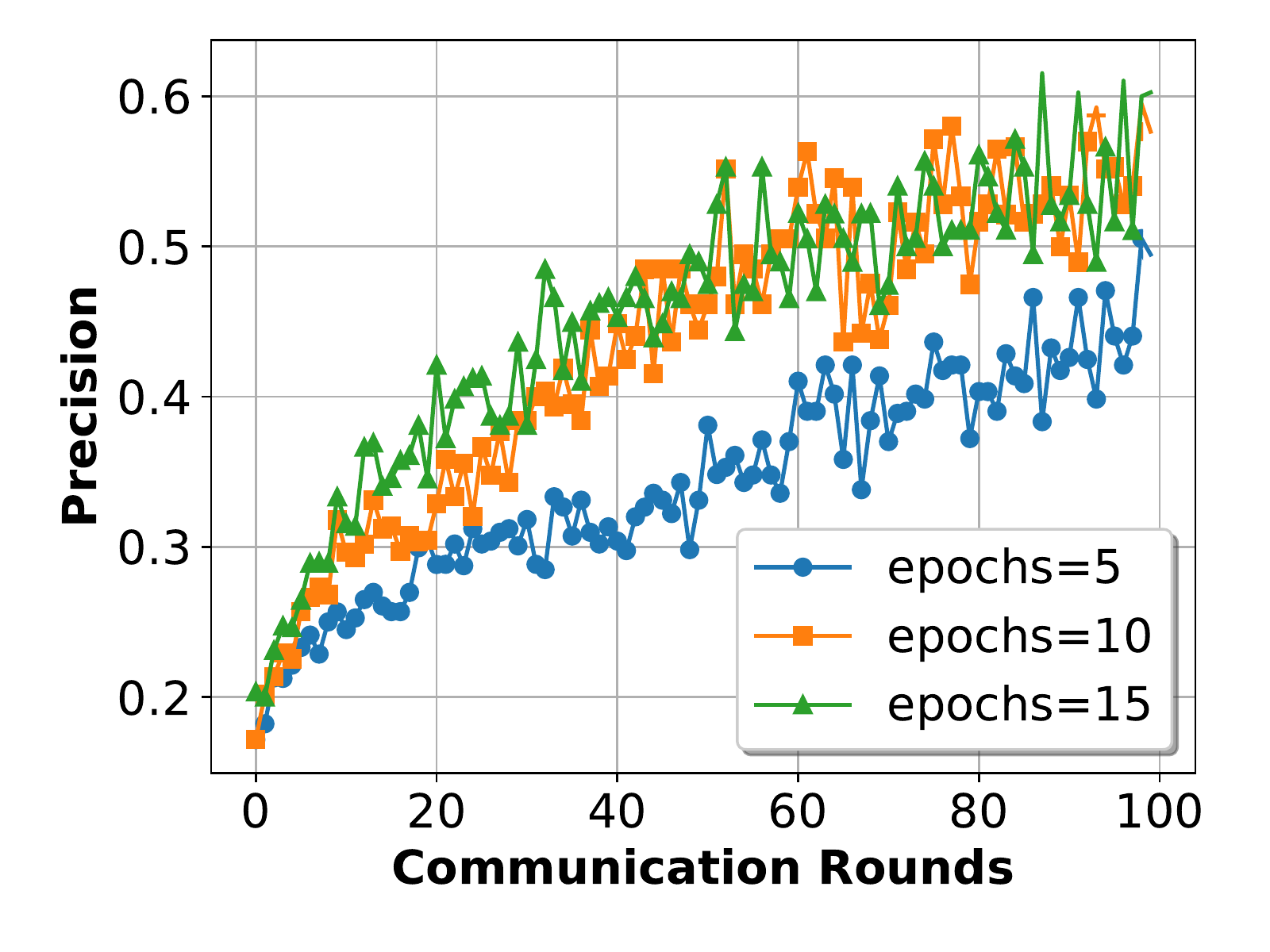}
         \caption{Precision}
         \label{fig:pre-epoch}
     \end{subfigure}
        \caption{Performance evaluation in terms of test accuracy, loss, F1 Score, and precision under various numbers of local epochs.}
        \label{fig:impact_of_epochs}
\end{figure*}

\section{Performance Evaluation}
In this section, we describe our system configurations and analyze the performance of $\mathsf{BDSP}$ system in various scenarios. 

\subsection{System Settings}
To demonstrate the applicability of our proposed system, we implement a proof-of-concept for the sharing model in a Blockchain-based network. We procedure the experiments in a server with Intel(R) Xeon(R) Gold 6248 CPU @ 2.50GHz. Table~\ref{tab:parameter-setting} presents the configurations of our experiments.

\subsubsection{Blockchain network} 
%
We implement Hyperledger Fabric (HF) for simulating a Blockchain network \cite{androulaki2018hyperledger}. HF was constructed by The Linux Foundation\footnote{\url{https://www.linuxfoundation.org/}} in 2015. It is an open-source and permissioned Blockchain framework. HF offers unique identity management and access control features. Hence, it is suitable for various industrial applications, such as banking, finance assests \cite{ma2019privacy}, and supply chains \cite{nguyen2020trusted}.

\subsubsection{Datasets} 
%
To conduct our experiments, we use a real credit card fraud dataset \cite{CreditCa44:online}. The dataset includes anonymous credit card transactions made by European cardholders. It contains $284,807$ credit card transactions in September $2013$ that happened over two days. Among the $284,807$ transactions, only $492$ ($0.17\%$) are fraudulent, making the dataset heavily skewed. Further, it contains $30$ features, of which only two are known, such as the amount and time of the transactions. 

\subsubsection{Evaluation metrics} To evaluate the performance of $\mathsf{BDSP}$, we employ various evaluation metrics, such as accuracy, test loss, F1 score, and precision. 
The accuracy metric is not the best suitable metric to use when evaluating imbalanced datasets as it can be misleading. Besides accuracy, we consider for performance metrics to evaluate the performance of $\mathsf{BDSP}$, namely test loss, F1 score, and precision. 
In addition, the communication overhead is also considered as a performance metric.

\subsection{Results}

To minimize the impact of over-fitting, we divide the data in the ratio of $8$:$2$ for training data and testing data, respectively. 

\subsubsection{Impact of the number of epochs} 
%
In FL algorithms, for each client, the number of epochs impacts the learning progress before we update the global model. Locally, more epochs will result in more progress in each training round, leading to a much faster convergence rate per round. Figure~\ref{fig:impact_of_epochs} demonstrates the performance of $\mathsf{BDSP}$ with a different number of local training epochs ranging from $5$ to $15$. 

The results show that we achieve better accuracy and the F1 score when increasing the number of local epochs. In particular, Figure~\ref{fig:acc-epoch} demonstrate that with epochs $=\{10,15\}$, $\mathsf{BDSP}$ achieves higher accuracy around $0.997$ compared to $0.996$ of epoch=$\{5\}$. The same observations are for the F1 score and precision as shown in Figure \ref{fig:f1-epochs} and Figure \ref{fig:pre-epoch}. 

\subsubsection{Impact of batch size} 
\begin{figure}[t!]
     \centering
     \begin{subfigure}[b]{0.35\textwidth}
         \centering
         \includegraphics[width=\textwidth]{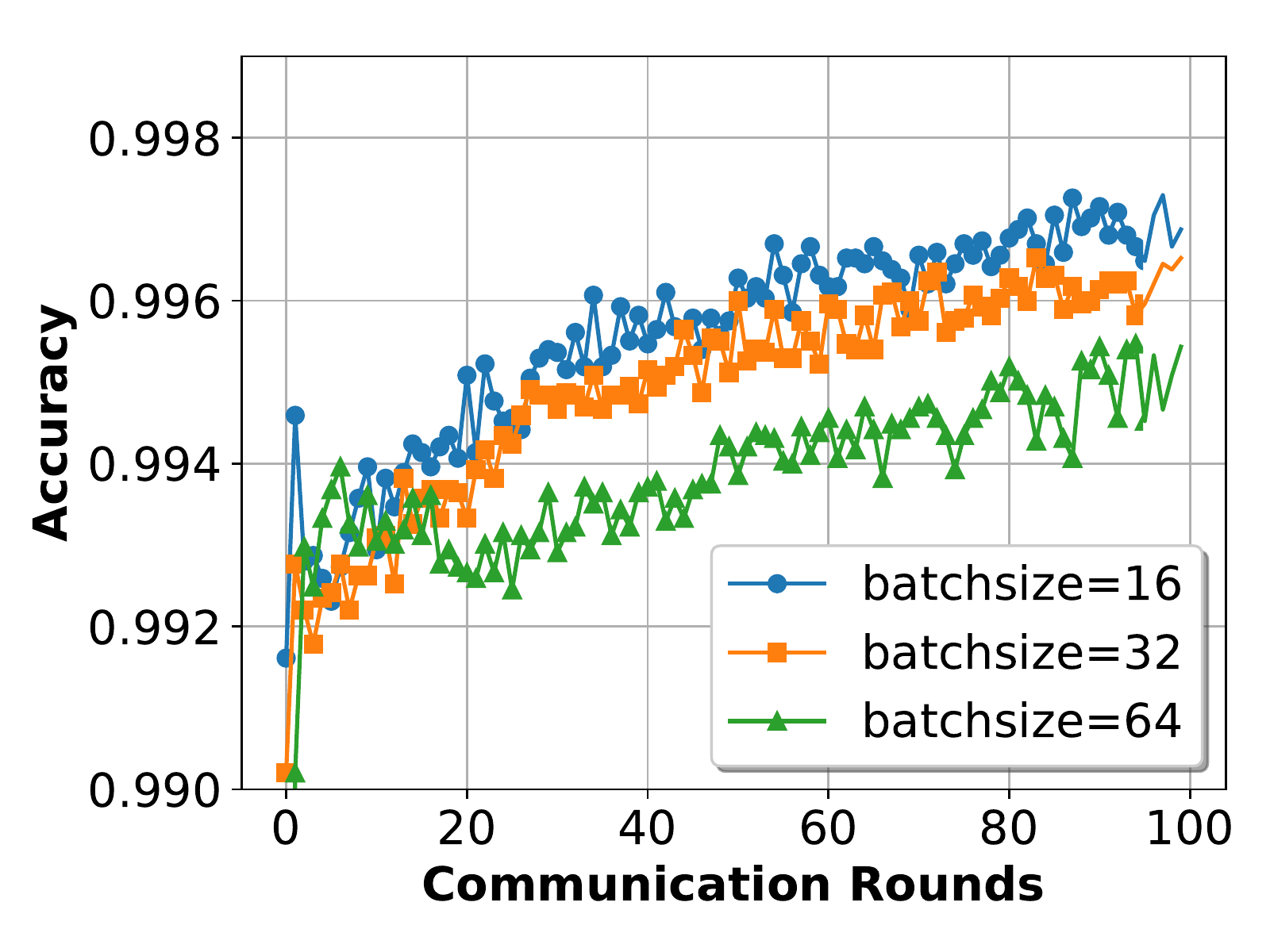}
         \caption{Test Accuracy}
         \label{fig:batch_acc}
     \end{subfigure}
    
     \begin{subfigure}[b]{0.35\textwidth}
         \centering
         \includegraphics[width=\textwidth]{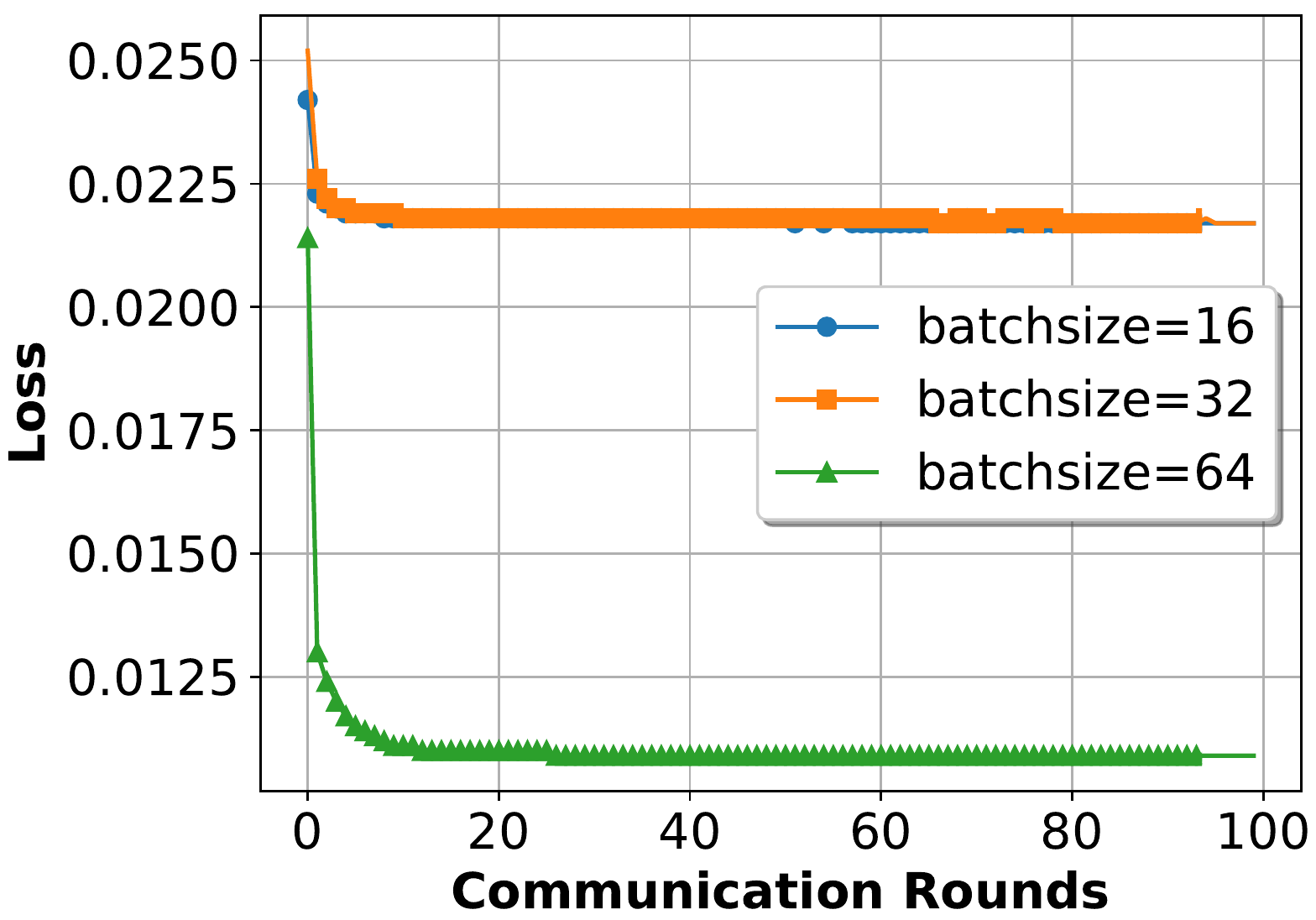}
         \caption{Test Loss}
         \label{fig:batch_loss}
     \end{subfigure}

     \begin{subfigure}[b]{0.35\textwidth}
         \centering
         \includegraphics[width=\textwidth]{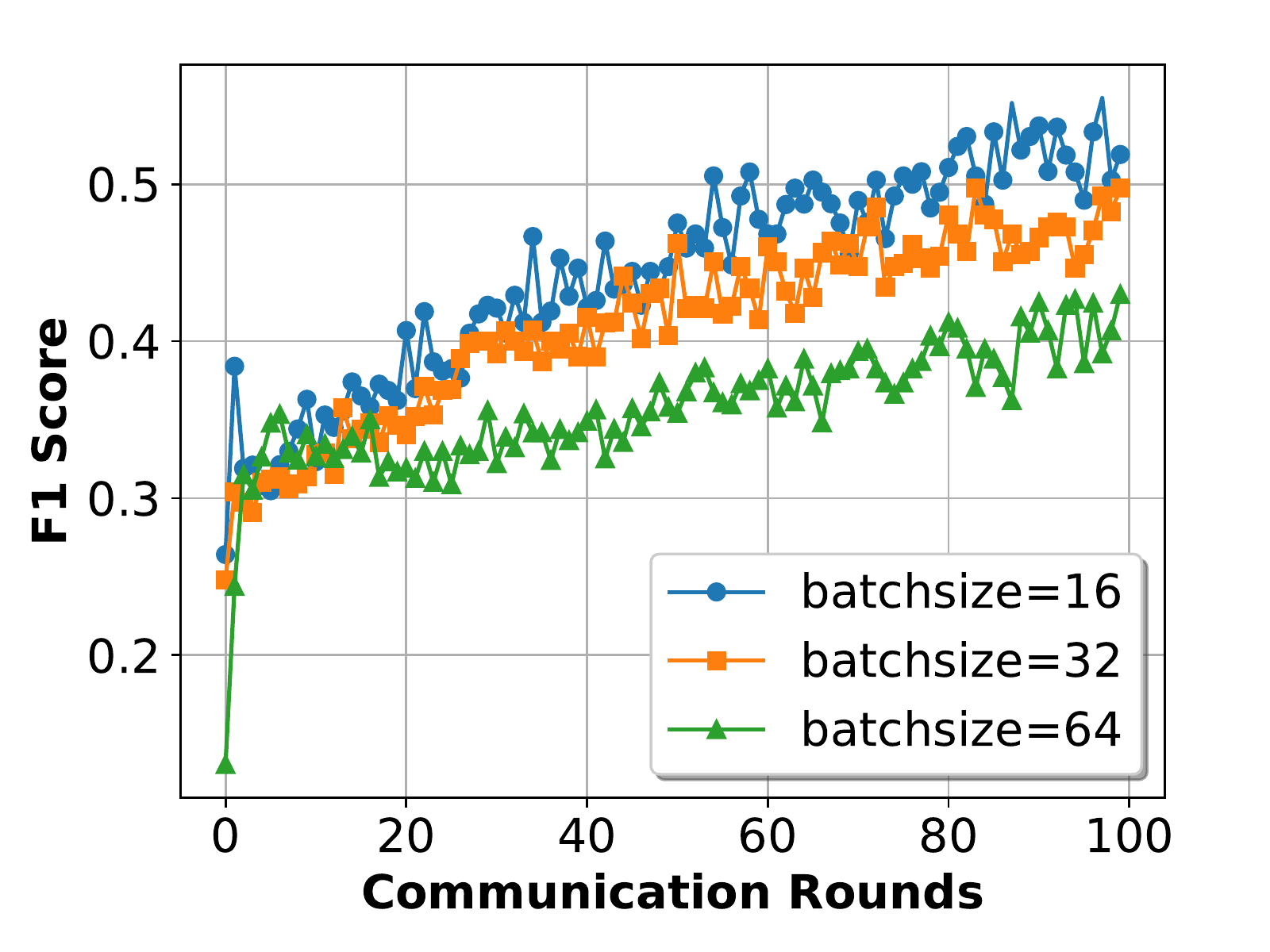}
         \caption{F1 Score}
         \label{fig:batch_f1}
     \end{subfigure}

     \begin{subfigure}[b]{0.35\textwidth}
         \centering
         \includegraphics[width=\textwidth]{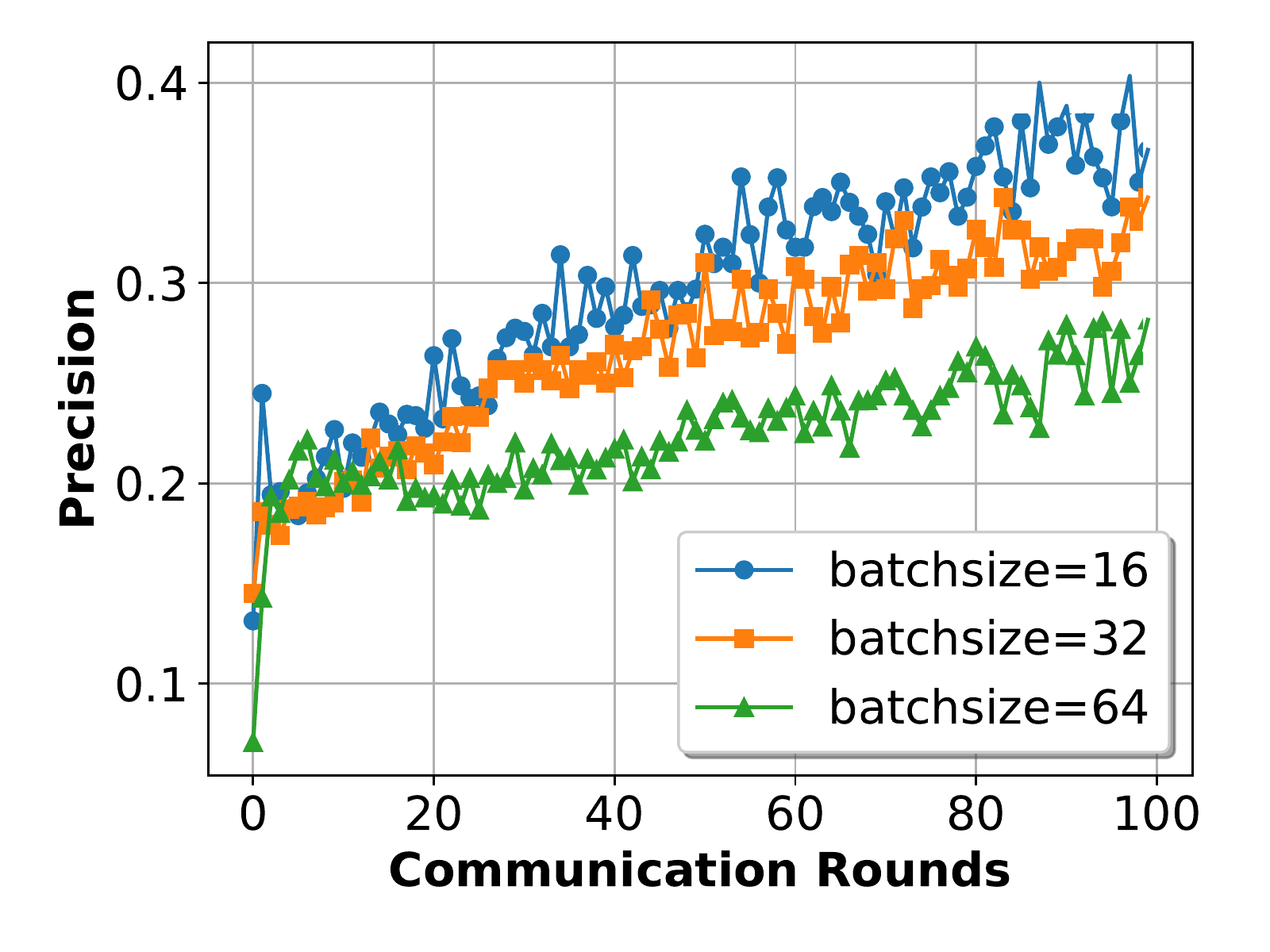}
         \caption{Precision}
         \label{fig:batch_pre}
     \end{subfigure}
        \caption{Impact of batch size on the performance evaluation.}
        \label{fig:impact_of_batchs}
\end{figure}

The batch size performs local updates in each interaction, significantly impacting the performance and network efficiency.
Figure~\ref{fig:impact_of_batchs} presents the performance of $\mathsf{BDSP}$ in varying batch sizes from $16$ to $64$. 
%
%
In these measurements, we fix the number of epochs at 10. Figure~\ref{fig:batch_acc} clearly shows the impact of batch size. When the batch size is $64$, our system achieves an accuracy of 0.994, which is lower than when the batch size is equal to $32$ or $16$. 

We see a similar observation for the F1 score and precision in Figures~\ref{fig:batch_f1} and~\ref{fig:batch_pre}, respectively. Increasing the batch size leads to lower performance in our system. However, we can avoid this problem by adjusting the learning rate during our training process. 

In addition, the loss measurement of $\mathsf{BDSP}$ in various numbers of epochs and batch sizes are demonstrated in Fig.~\ref{fig:loss-epoch} and Fig.~\ref{fig:batch_loss}, respectively. The fact is that the number of epochs and batch size significantly impact the performance of training models. For example, a larger amount of batch size can handle more requests but increase the delay of the training process. 

\begin{figure}[!b]
    \centering
    \includegraphics[width=0.7\linewidth]{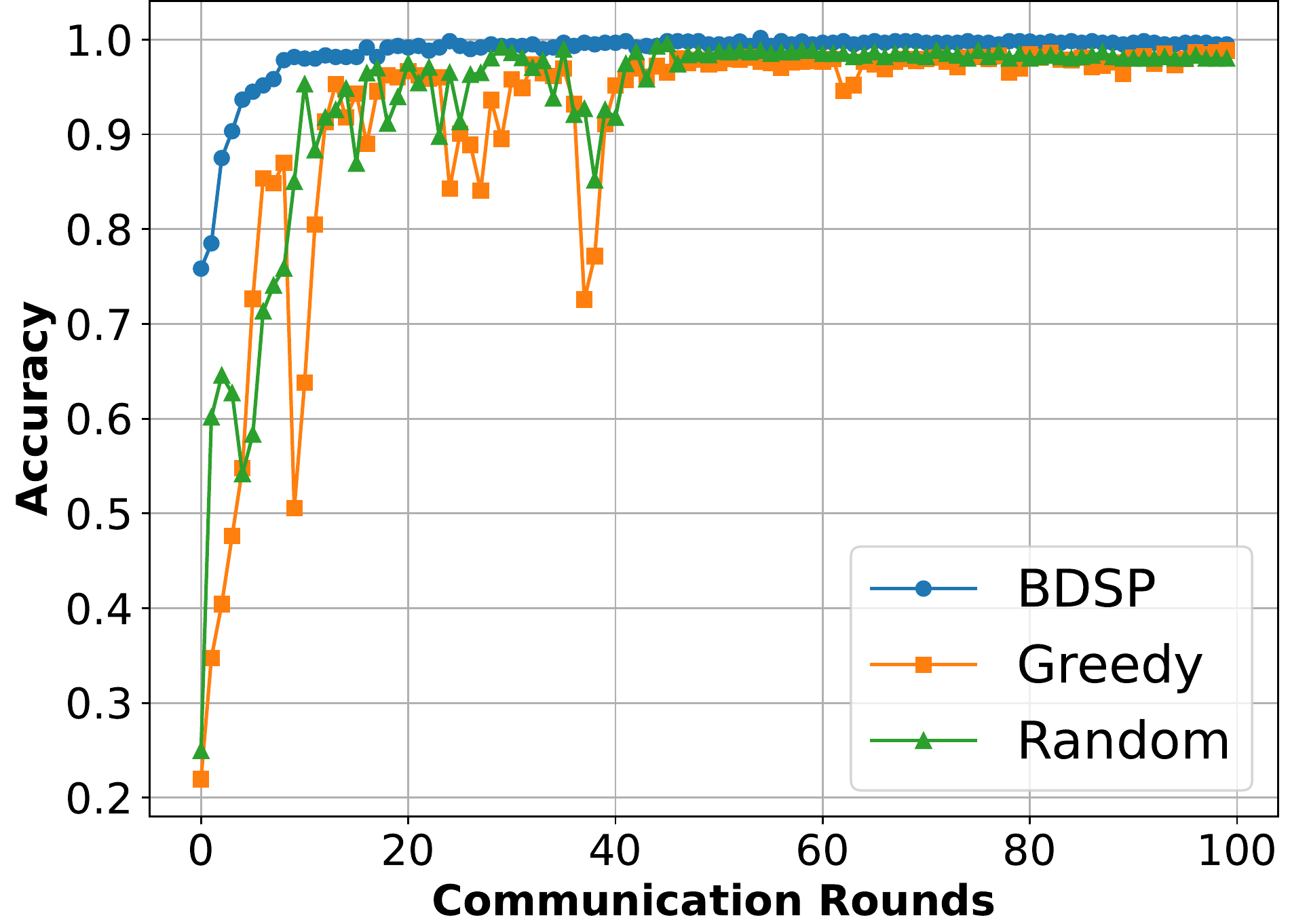}
    \caption{Average Accuracy of different organizations}
    \label{fig:organization_level}
\end{figure}

\subsubsection{Average Accuracy at organization-level} In Fig. \ref{fig:organization_level}, we analyze and compare the average accuracy at the organization level as the performance of the trained model on their local dataset.
We compare our $\mathsf{BDSP}$ against two intuitive baselines: i) greedy algorithm \cite{balakrishnan2021diverse} in which the aggregation server collects the local gradients of all clients and chooses a subset of clients providing the greatest marginal gain, and ii) random sampling strategies which choose clients randomly. 
We can see that $\mathsf{BDSP}$ enhances improvement at the organization-level performance as around $6.2\%$ and $5.1\%$ higher than the greedy method and random sampling, respectively. The reason is that the appropriate of each local update is collected and aggregated in the global model following the contribution of the selected organization in every global training round.

\subsubsection{Communication efficiency}
%
%
In addition, we measure the average time to achieve of $90\%$ of the accuracy of $\mathsf{BDSP}$ in comparison with these two baselines. The results show that $\mathsf{BDSP}$ requires around $2\pm 1$ rounds to achieve $90\%$, while the greedy method and random sampling require around $9\pm 1$ and $5\pm 1$ to achieve it, respectively. This demonstrates that $\mathsf{BDSP}$ improves the communication and computation overhead. 

In $\mathsf{BDSP}$ design, we aim to optimize the communication overhead between various organizations during the training process and the storage overhead on the Hyperledger network. Therefore, we store the local updates from organizations to the off-chain IPFS storage, and we only record the hashed version of SHA2-256 of the local updates sized at $256$ bits on-chain. Meanwhile, the size of a local update in our settings is around $56$ bytes; with the increasing number of organizations, the number of local updates recorded on the chain will be enormous, so it significantly increases the size of the ledger on-chain.

\section{Conclusion}
%
In this paper, we study the strategies for the collaborative sharing of data among organizations in order to guarantee data privacy. We first analyze the current privacy-preserving techniques that can support sharing data and provide our observations as the benchmark for selection. Then, we propose a Blockchain-based data-sharing framework integrated with federated learning called $\mathsf{BDSP}$. 
Next, we study the problem of selected organizations for the training process based on the contribution of each organization, and then we address the unbalance and lack of datasets in enterprise organizations and accordingly design a mechanism to rebalance the datasets among organizations for better training performance. 
In addition, we provide a mechanism to evaluate the data valuation of training datasets from organizations, which can be used to calculate the incentive or contribution of organizations in a joint game. We prove that our proposed solution improves the training accuracy of organizations and reduces the communication overhead. 
For future works, we are going to emphasize the practical deployment of multiple sources of datasets in different types and formats. Moreover, sharing data over cross-chain networks could be addressed in future research.


\normalem
\bibliographystyle{IEEEtran}
\bibliography{ref}

\end{document}